\documentclass[10pt,journal,compsoc]{IEEEtran}

\ifCLASSOPTIONcompsoc
  \usepackage[nocompress]{cite}
\else
  \usepackage{cite}
\fi

\ifCLASSINFOpdf
   \usepackage[pdftex]{graphicx}
\else
\fi

\usepackage{underscore}
\usepackage[document]{ragged2e}
\usepackage[ruled,vlined]{algorithm2e}
\usepackage{dblfloatfix}
\usepackage{caption}
\usepackage{hyperref}
\usepackage[all]{hypcap}
\usepackage[T1]{fontenc}
\DeclareUnicodeCharacter{0301}{\'{e}}

\usepackage{scalerel}
\usepackage{tikz}
\usetikzlibrary{svg.path}

\definecolor{orcidlogocol}{HTML}{A6CE39}
\tikzset{
  orcidlogo/.pic={
    \fill[orcidlogocol] svg{M256,128c0,70.7-57.3,128-128,128C57.3,256,0,198.7,0,128C0,57.3,57.3,0,128,0C198.7,0,256,57.3,256,128z};
    \fill[white] svg{M86.3,186.2H70.9V79.1h15.4v48.4V186.2z}
                 svg{M108.9,79.1h41.6c39.6,0,57,28.3,57,53.6c0,27.5-21.5,53.6-56.8,53.6h-41.8V79.1z M124.3,172.4h24.5c34.9,0,42.9-26.5,42.9-39.7c0-21.5-13.7-39.7-43.7-39.7h-23.7V172.4z}
                 svg{M88.7,56.8c0,5.5-4.5,10.1-10.1,10.1c-5.6,0-10.1-4.6-10.1-10.1c0-5.6,4.5-10.1,10.1-10.1C84.2,46.7,88.7,51.3,88.7,56.8z};
  }
}

\newcommand\orcidicon[1]{\href{https://orcid.org/#1}{\mbox{\scalerel*{
\begin{tikzpicture}[yscale=-1,transform shape]
\pic{orcidlogo};
\end{tikzpicture}
}{|}}}}

\begin{document}
\title{Network Participation and Accessibility of Proof-of-Stake (PoS) Blockchains: A Cross-platform Comparative Analysis}
\author{Jiseong~Noh \orcidicon{0009-0001-3951-9809}, Donghwan~Kwon \orcidicon{0009-0005-4097-9465}, Soohwan~Cho \orcidicon{0000-0002-2956-0082} and Neo~C.K.~Yiu \orcidicon{0000-0002-9047-5207},~\IEEEmembership{Member,~IEEE}
\IEEEcompsocitemizethanks{
\IEEEcompsocthanksitem Jiseong Noh is with Klaytn Foundation and TEEware, E-mail: joseph.noh@klaytn.foundation and jsnoh@teeware.kr
\IEEEcompsocthanksitem Donghwan Kwon is with Klaytn Foundation, E-mail: aidan.kwon@klaytn.foundation
\IEEEcompsocthanksitem Soohwan Cho is with Klaytn Foundation, E-mail: iron.cho@klaytn.foundation
\IEEEcompsocthanksitem Neo C.K. Yiu is with Klaytn Foundation and Department of Engineering, University of Cambridge, U.K., E-mail: neo.yiu@klaytn.foundation and cky26@cam.ac.uk}
\thanks{Manuscript first submitted for preprint on May 16, 2023.}}

\IEEEtitleabstractindextext{
\justify
\begin{abstract}
The comparative analysis examined eleven Proof-of-Stake (PoS) consensus-based blockchain networks to assess their openness based on five indicative metrics. These metrics include those of decentralization-related aspects, such as the number of validators and capital concentration, and participation-related aspects, including entry capital requirements and economic network stability. This is to assess and characterize the openness of Proof-of-Stake blockchain networks. The analysis suggested that networks with higher openness included Solana and Avalanche, while BNB Chain, Klaytn, and Polygon measured with lower levels of openness. According to the comparative analysis, Ethereum scored high on network openness in terms of the number of participants and the cost of running the chain, but scored relatively low on capital concentration and staking ratio, which is likely due to the low ratio of staked ether (ETH) to circulating supply and the significant stakes in staking pools like Lido. Permissioned blockchains such as Klaytn and Polygon have limited openness, which suggests the need to take the level of openness into account when transitioning into a permissionless blockchain architecture with a more decentralized setting.
\end{abstract}

}

\maketitle
\IEEEdisplaynontitleabstractindextext
\justifying
\IEEEpeerreviewmaketitle

\ifCLASSOPTIONcompsoc
\IEEEraisesectionheading{\section{Introduction}\label{sec:introduction}}
\else
\section{Introduction}
\label{sec:introduction}
\fi

\IEEEPARstart{I}{n} this article, the openness levels of different blockchain networks running on a Proof of Stake (PoS) \cite{8}\cite{9} consensus basis are analyzed, identifying their strengths and weaknesses, and suggesting ways to increase their openness.

Proof-of-stake is a method of creating blocks and achieving consensus based on the number of assets staked. Typically, it uses a consensus method proportional to the stake (therefore the “Proof of Stake”), which means that validators with a larger stake have more influence on block creation and consensus process. The problem with this approach is that some validators with a large stake can become more powerful. To mitigate this, some blockchain networks limit the usage of stake to validator qualifications and achieve consensus in proportion to the number of validators. Once you become a validator on such a network, you can participate in the consensus process on the same footing as other validators, regardless of the size of your stake on the blockchain network. 

The openness of a blockchain is related to decentralization to some extent, but not exactly the same \cite{25}. First of all, openness focuses on the accessibility to blockchain network participation. Considering whether anyone can join as a validator, create and verify blocks, and how much each network participant can contribute to be important, it also looks at how reliably a network can be operated and maintained against potential security attacks or any malicious actions from the network participants. These factors help measure the openness of a blockchain network.

Meanwhile, decentralization focuses on how evenly functions, control and information could be distributed amongst participants on a blockchain network. Openness, in other words, related to a concept of decentralization but has a broader meaning.

While there has yet been a standard for assessing the openness of a blockchain, we can look into decentralization factors such as the number of network participants, the number of meaningful participants, accessibility, sustainability and network security from an economic perspective. In this article, we would like to compare the openness levels of blockchains based on the following \underline{\emph{seven}} metrics.

\begin{enumerate}
  \item \emph{The number of validators}: The number of validators refers to the number of nodes that directly participate in a blockchain network and create and validate blocks, and it can serve as one of the important indicators of network openness. In general, the number of validators on an open blockchain is higher than that of validators on a permissioned blockchain.
  \item \emph{The capital required for participation}: The capital required to participate is closely related to the openness of a blockchain network. Networks with lower capital requirements for participation become more economically accessible, and this can help broaden the base of users participating in running validators.
  \item \emph{Capital concentration}: A blockchain with an even distribution of staked capital can maintain high security through a consensus with a large number of validators. Proof-of-Stake blockchains often determine consensus in proportion to staked capital. If too many stakes are clustered in particular validators, a block can be created with the consensus of only a few validators, which is not appropriate for an open network. Therefore, we could conclude that the more evenly staked capital is distributed, the more open the blockchain is. 
  \item \emph{Operating costs}: The lower the cost of running validators, the more users can be encouraged to participate as validators on an ongoing basis.
  \item \emph{Network stability from an economic perspective}: The economic stability of a blockchain network plays an important role in protecting the network from external attackers. There are two factors to consider for network stability;
  \item \emph{Staking ratio}: The higher the percentage of staked assets in circulation, the more stable the network. A higher staking ratio can make it harder for attackers to acquire the native tokens they need to disrupt a network. 
  \item \emph{Cost of attack}: The cost for an attacker to compromise a network is an important factor for network stability. The higher the cost of an attack, the more economically disadvantageous it is to attack the network, which may decrease the chance for the attacker to make such an attempt.
\end{enumerate}

This paper is a cross-platform comparative analysis and systemization of knowledge in decentralization, participation and economic aspects of existing permissionless and permissioned blockchains in the blockchain industry, it covers the analysis of nine permissionless blockchain networks \cite{16}\cite{20} and two permissioned blockchain networks that plan to gradually transform to become permissionless and measure and analyze the levels of openness respectively. This comparative study will first look at the related work in measuring the levels of openness and its participation-related aspects. It will then cover the mechanism and operation of measuring the openness level of different blockchains and its summary concluding from the comparative analysis with the suggested future work for improvement in levels of openness and participation of different blockchain protocols.

\section{Related Work}
Several studies have been conducted to evaluate and compare different blockchain platforms. \cite{1} investigates how to measure the decentralization of blockchain networks, while \cite{2} compares different blockchain platforms in the context of their ecosystems. In a similar vein, \cite{3} performs an analysis of different blockchain platforms with a focus on comparing consensus protocols. Furthermore, \cite{4} argues that public blockchain networks are becoming increasingly centralized. Finally, \cite{5} provides a comprehensive summary of research work in blockchain decentralization, highlighting the use of a decentralization index to measure the level of decentralization of public blockchains.

Several studies have investigated various aspects of Proof of Stake networks. \cite{6}, \cite{10}, \cite{11}, \cite{14}, and \cite{15} have focused on wealth distribution in different blockchain networks and linked it to potential security threats due to asset centralization. Proof of Stake and its variants have been described and analyzed in \cite{8} and \cite{9}. Many studies have investigated various attack scenarios against Proof of Stake networks: \cite{12} argues that a 51\% attack on Proof of Stake is unlikely, and \cite{13} investigates double-spend attacks on Proof of Stake chains and measures their probability. \cite{17} identifies attack vectors against the consensus algorithm of Proof-of-Stake and compares several Proof-of-Stake blockchain networks in terms of their resistance to these attacks. \cite{19} determines the economic value required to launch a 51\% attack on a blockchain, and \cite{23} provides insights into the Sybil attack and the double-spend attack.

In the realm of permissionless blockchains, \cite{16} suggests that permissionless blockchains have more potential than permissioned blockchains. \cite{20} elaborates on the differences between permissioned and permissionless blockchains while \cite{22} elaborates on the performance and cost evaluation of public blockchains with a case study on NFT marketplace applications. Regarding the openness of blockchain networks, \cite{19} argues that the Nakamoto coefficient can be utilized as a metric to measure the decentralization of blockchain networks.

\section{Measuring Openness Levels}
\subsection{The Number of Validators}
First, let's look at the number of validators: the more validators a network has, the more open it is, as the opinions of more participants can be taken into account. In general, we may expect that an open blockchain network would have a higher number of validators than a permissioned blockchain network because it is structured to be open to anyone.

\begin{table}[h]
    \centering
    \captionsetup{justification=centering}
    \includegraphics[width=0.5\textwidth]{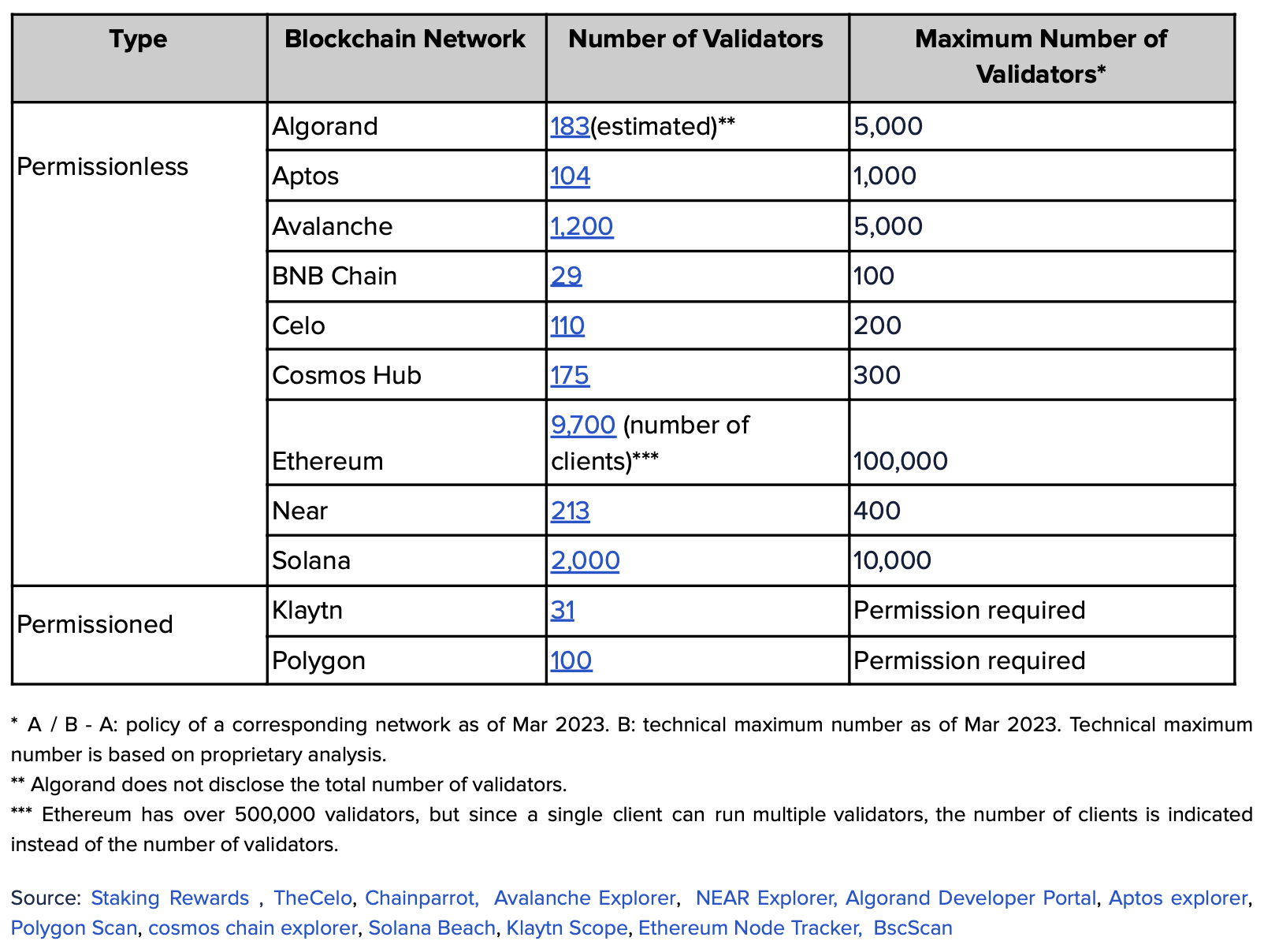}
    \caption{\textit{Number of validators by blockchain network as of March 2023}}
    \label{table:1}
\end{table}

For instance, Bitcoin, one of the open blockchains, has about \href{https://bitnodes.io/}{17,000} nodes. In contrast, EOS, one of the permissioned blockchains, has only \href{https://eosauthority.com/}{21} validators.

The difference in the number of validators between permissionless and permissioned blockchains is summarized in \textit{Table~\ref{table:1}}. It also depicts that the number of validators varies among open blockchains. This is due to different factors such as conditions for validator participation and the maximum limit of validators.

\begin{figure}[h]
    \centering
    \captionsetup{justification=centering}
    \includegraphics[width=0.5\textwidth]{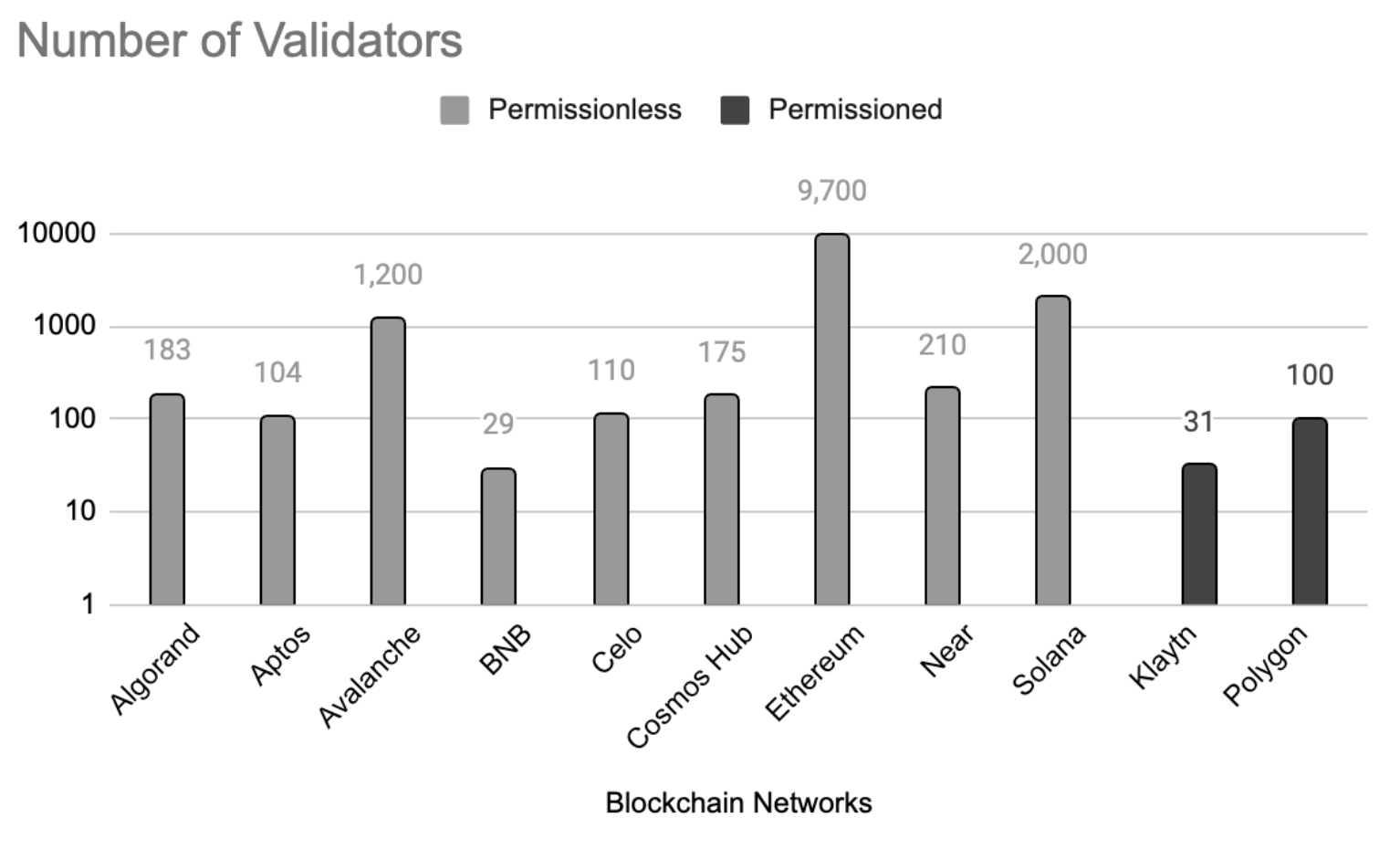}
    \caption{\textit{Number of validators on different blockchain networks}}
    \label{fig:1}
\end{figure}

According to the summary on the number of validators on different blockchain networks as demonstrated in \textit{Fig.~\ref{fig:1}}, BNB Chain is a permissionless blockchain, but it limits the maximum number of validators to 29. Like BNB Chain, many open blockchains adopt a strategy of limiting the maximum number of validators to increase reliability, even if it means restricting some openness.

For blockchains with multiple validator nodes operating with a single client, such as Ethereum, it is important to consider whether such multiple validators should be considered as multiple objects or one object when assessing their openness. For the purposes of this article, all validators running on a single client are considered one validator because if a client is down for some reason, all validators connected to the client could thus be shut down simultaneously. Therefore, this analysis article indicates the number of Ethereum clients instead of the number of its validators.

\subsection{Capital Requirement for Participation}
The capital required to participate has a significant impact on the openness of a blockchain. The lower the initial capital cost, the lower the barrier to entry, giving more network users the opportunity to participate as validators, which would enhance openness. However, lower initial capital requirements can also lead to the enhanced risk of a malicious act. One of the key approaches to prohibit such malicious behaviors could be "slashing", but a lack of capital to slash could weaken the effectiveness of the restriction and allow malicious actors to try committing an attack with a lower opportunity cost.

\begin{table}[h]
    \centering
    \captionsetup{justification=centering}
    \includegraphics[width=0.5\textwidth]{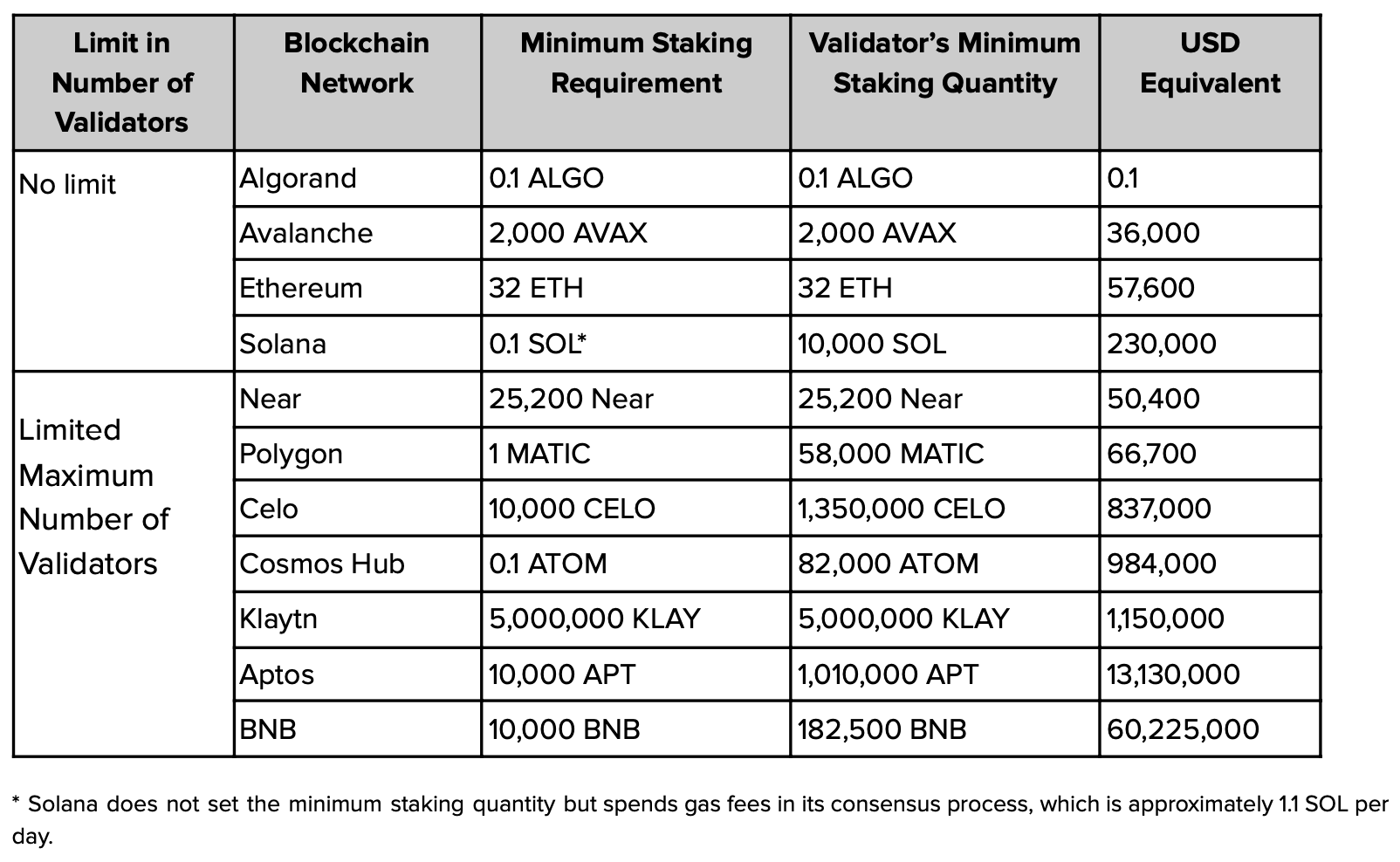}
    \caption{\textit{Minimum capital to be qualified as a validator as of the end of March 2023}}
    \label{table:2}
\end{table}

\textit{Table~\ref{table:2}} indicates that each blockchain network has a different minimum capital requirement to serve as a validator, which provides some insights into the openness levels of blockchain networks.

BNB Chain has a relatively high minimum capital requirement, which suggests that it values trust toward selected operators more than openness compared to other blockchains. Neither Cosmos Hub nor Algorand has a minimum staking requirement, so they can be considered more open in terms of minimum staking requirements. While Solana has no minimum capital requirement to participate as a validator, but it has a unique structure that should consider gas fees incurred from its consensus process. With the adoption of this strategy, Solana has removed the barrier to entry of having to initially stake a large amount of capital.

The amount of capital staked as a collateral by validators with the least capital among the validators participating in each network is summarized in \textit{Table~\ref{table:2}}. This suggests an idea of how much capital is required to participate in a blockchain network as a validator. 

To become a validator on the blockchain networks such as BNB Chain and Aptos, one is required to rank in the top validators, which requires much more capital than the minimum staking requirement. In other words, one would need to stake a much larger amount of capital than the minimum staking requirement to participate in a blockchain network as a validator. Ordinary users could find it difficult to become validators on the blockchain networks with such demanding capital requirements, hence they may need help and support from blockchain foundation of the blockchain networks and other related organizations.

Blockchain networks such as Avalanche, Ethereum, and Solana do not limit the number of validators, so anyone could participate in the blockchain network as a validator as long as they meet the minimum staking requirements. These blockchain networks have a large number of validators with a relatively small staking amount. This allows more users to serve in the blockchain network as validators and helps increase the openness of the network itself. Ethereum has set both the minimum and maximum staking amount equally at 32 ETH. Ethereum has a large number of validators on the main-net because those well-funded groups operating multiple validator nodes, each of whom is staking 32 ETH.

\subsection{Capital Concentration}
The openness of a blockchain network is related to the distribution of staked capital. Particularly if stakes are concentrated on a few validators or parties participating in a blockchain network where a consensus is reached in proportion to staked capital, a particular validator may have too much influence over the consensus process. This reduces the openness of the network and may lead to centralization and control over the blockchain network \cite{6}, \cite{11}, \cite{14}, \cite{15}. The more evenly staked capital is distributed, the more balanced power and control exist amongst validators, and we could conclude that the blockchain network is highly open.  

For blockchain networks that use a consensus mechanism proportional to the number of validators, the influence is more evenly distributed among validators, and more validators are required to reach a consensus. Therefore, these networks are relatively more open in terms of capital concentration.

\begin{table}[h]
    \centering
    \captionsetup{justification=centering}
    \includegraphics[width=0.5\textwidth]{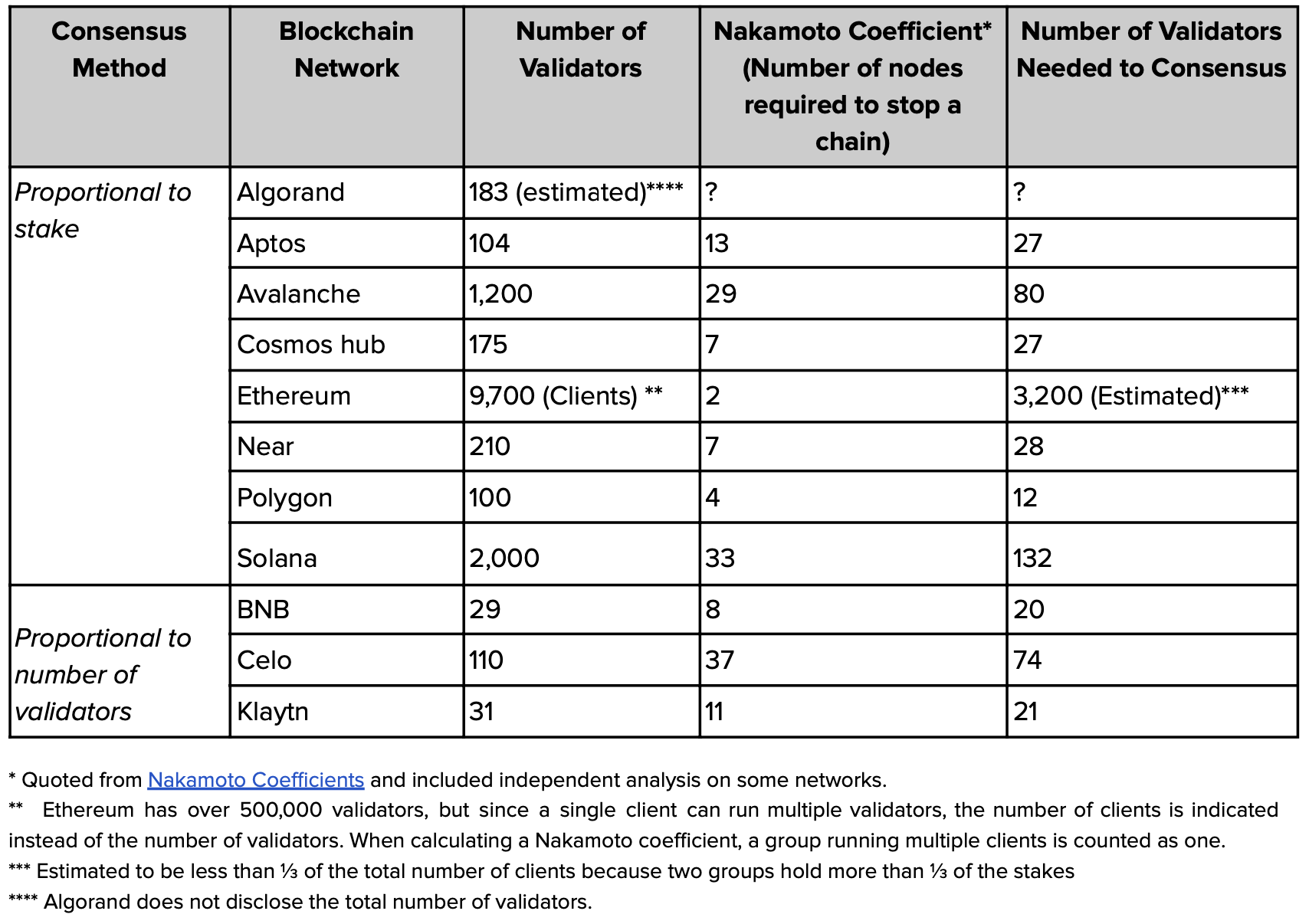}
    \caption{\textit{Number of validators with significant influence on consensus by blockchain network}}
    \label{table:3}
\end{table}

\begin{figure}[h]
    \centering
    \captionsetup{justification=centering}
    \includegraphics[width=0.5\textwidth]{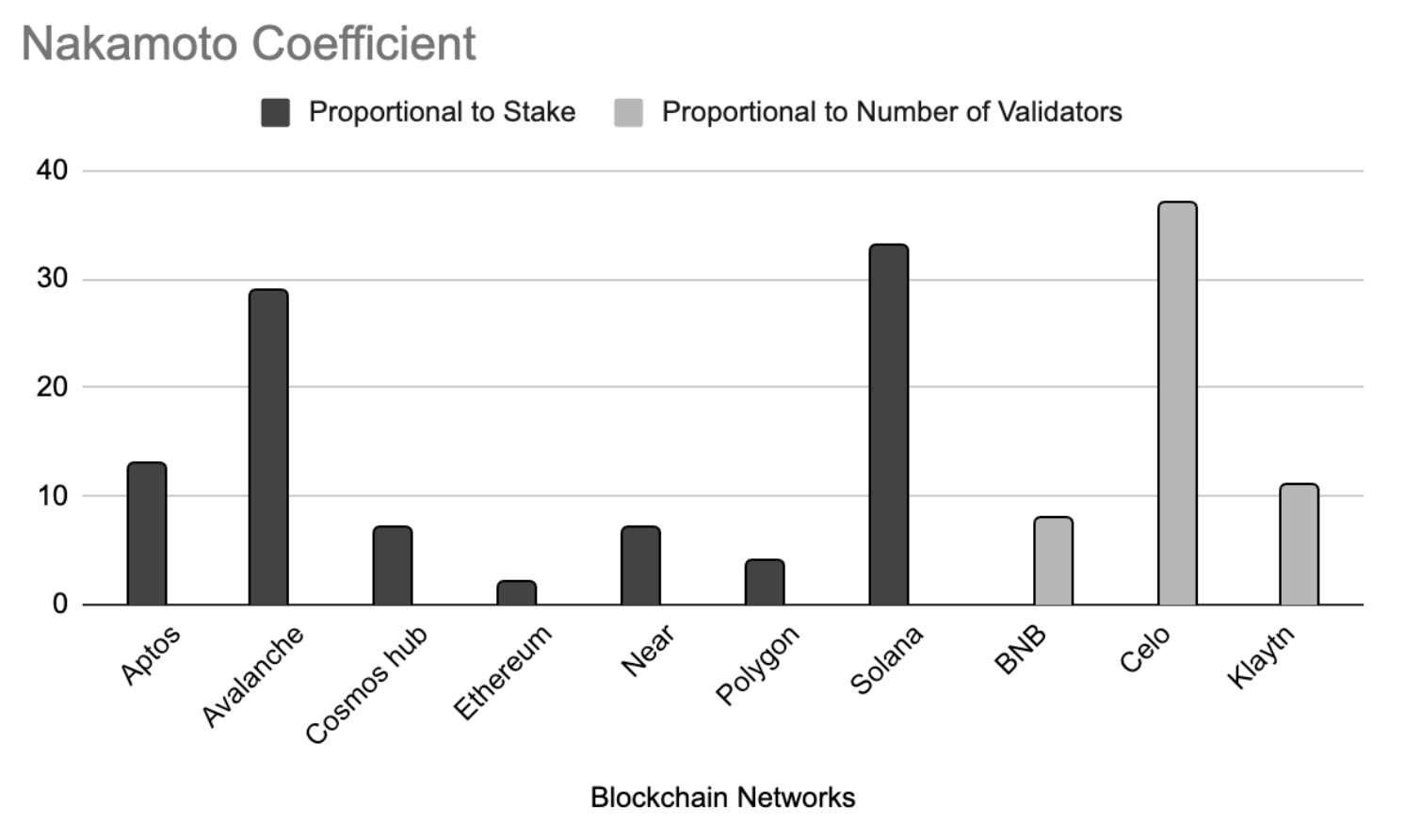}
    \caption{\textit{The corresponding Nakamoto Coefficient of different blockchain networks}}
    \label{fig:2}
\end{figure}

\textit{Table~\ref{table:3}} demonstrates that a blockchain network utilizing a consensus mechanism proportional to the number of validators has a relatively lower Nakamoto coefficient than that of a network using a consensus mechanism proportional to the number of validators, as listed in \textit{Fig.~\ref{fig:2}}. When a small number of validators have a large stake in a consensus mechanism proportional to their stake, they exercise a large influence on the network, and any problem with their behavior could compromise the network stability. For instance, a considerable amount of stake in Ethereum is in staking pools like Lido, which appears that more than 1/3 of the capital is concentrated on a few groups. On the other hand, blockchains that is based on a consensus mechanism proportional to the number of validators could provide a relatively higher level of decentralization with a smaller number of validators due to the fact that all validators have an equal amount of voting power.

When we look at the number of validators required to reach consensus, it does not vary much. While Solana has over 2,000 validators, the number of validators required to reach a consensus is around 130, or 7\% of the total. While a network which reaches a consensus proportional to the stake seems to have a large number of participating validators, the network could operate with the consensus of a small number of validators. However, Ethereum would need a sufficiently large number of clients for consensus.

\subsection{Operational Costs}
The cost of operating validator nodes could pose a significant impact on the openness of blockchain networks. Lower operating costs could enhance network openness by lowering barriers to participation. Higher operating costs could discourage participation by making it more expensive to retain validators. In addition, to offset operating costs, validators could face increased pressure to cash out the tokens they are rewarded with. This leads to an increased supply of the tokens, which could affect the token price.

\begin{table}[h]
    \centering
    \captionsetup{justification=centering}
    \includegraphics[width=0.5\textwidth]{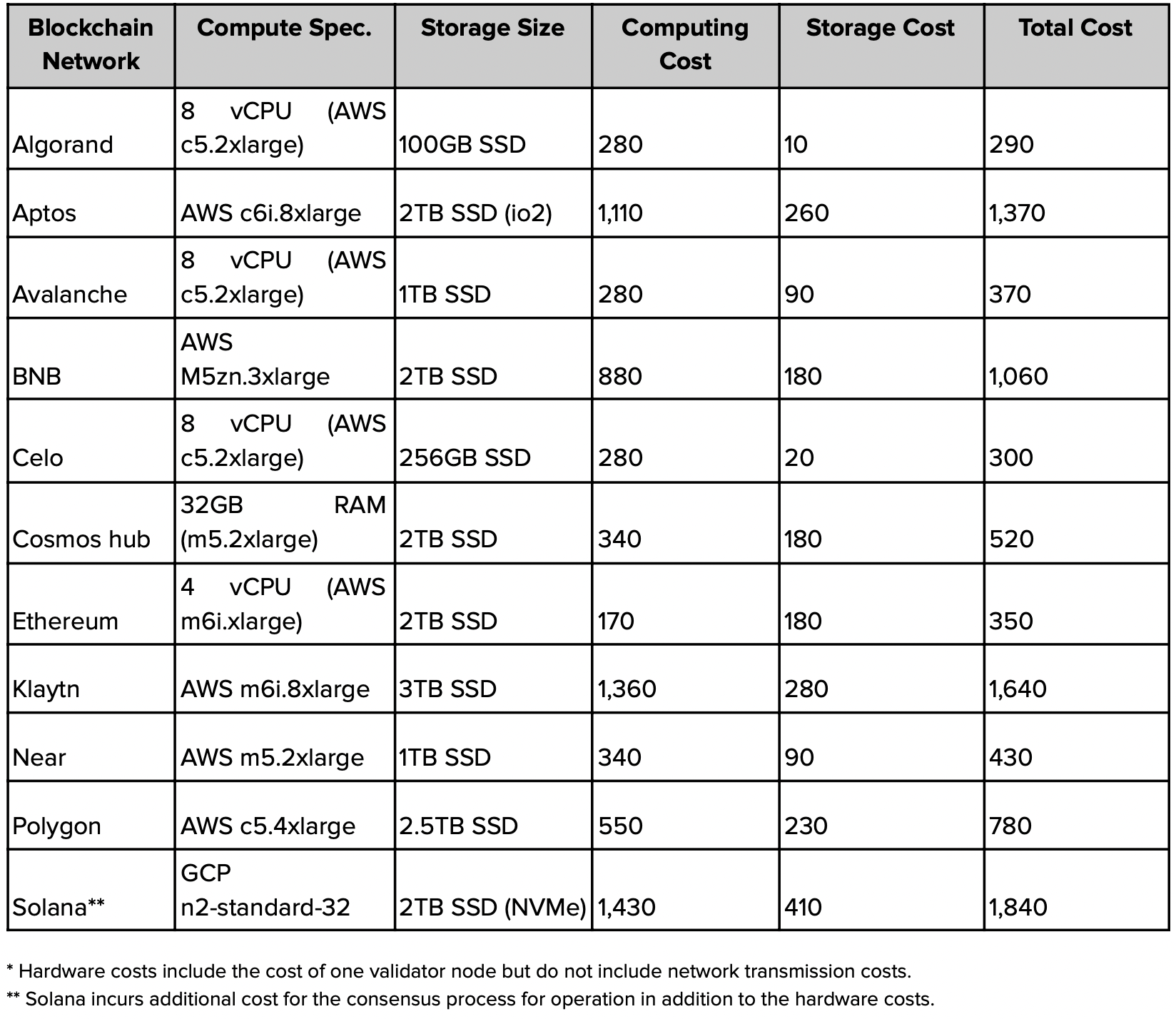}
    \caption{\textit{Validator minimum specifications and monthly hardware costs* (unit: USD)}}
    \label{table:4}
\end{table}

\textit{Table~\ref{table:4}} summarizes the monthly hardware costs based on AWS EC2 instances according to validator minimum specifications of different blockchain networks. The hardware specifications are quoted from each network's websites. According to \textit{Table~\ref{table:4}}, hardware specifications for validator operations vary by the blockchain network, so do the monthly hardware costs. This would enable a rough comparison of the operating costs to participate as a validator in the blockchain network.

\begin{figure}[h]
    \centering
    \captionsetup{justification=centering}
    \includegraphics[width=0.5\textwidth]{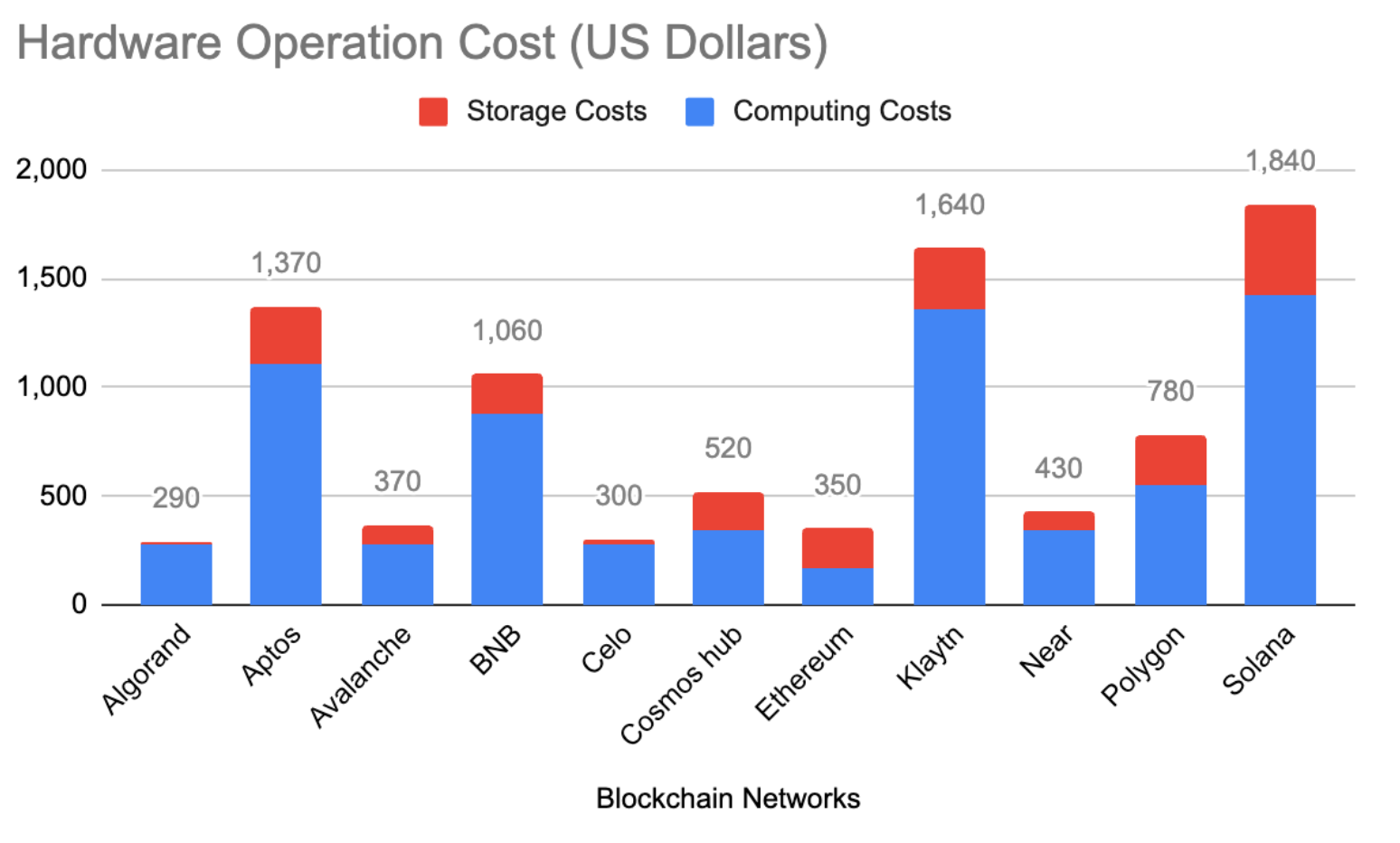}
    \caption{\textit{The hardware operation cost (in USD) of different blockchain networks}}
    \label{fig:3}
\end{figure}

As high-performance blockchain networks \cite{19} are optimized for high throughput and performance, their operating costs are relatively high. They require more CPU cores and memory as well as having a relatively high number of transactions per block, increasing the block size hence requiring more storage space; as such, the total operating costs could go up. \textit{Fig.~\ref{fig:3}} indicates that networks categorized as a high performance blockchain such as Aptos, Klaytn, and Solana have relatively high operating costs. This indicates that high-performance blockchains could be relatively weaker in terms of openness. In addition, high-performance blockchains in general have short block times and process many transactions per block, which means that a lot of data needs to be transferred and synchronized among validators in a short amount of time. Thus, the network transmission costs are also likely to be relatively higher.

If so, we can estimate how much upfront capital is needed to cover operating costs with the rewards for running a validator. If these costs are low and the rewards are sufficient, this could be an incentive to participate as a validator \cite{24}. Of course, this estimate is based on the reward level and a native token price at the moment, which can change at any time.

\begin{table}[h]
    \centering
    \captionsetup{justification=centering}
    \includegraphics[width=0.5\textwidth]{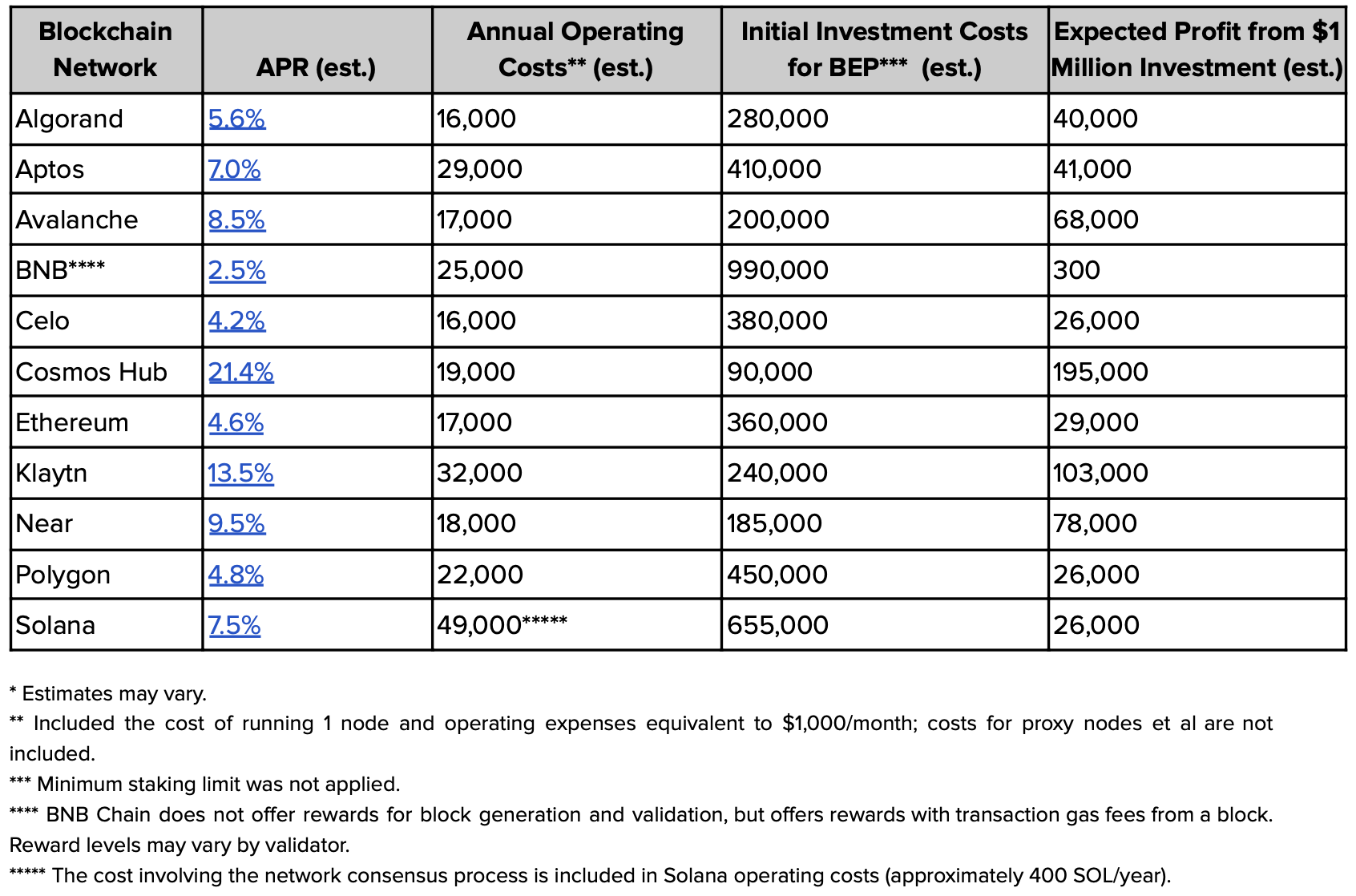}
    \caption{\textit{Initial investment required to cover the cost of running validators as of March 2023* (unit: USD)}}
    \label{table:5}
\end{table}

The estimated initial investment costs and annual reward rate for operating each network validator is summarized in \textit{Table~\ref{table:5}}. In this table, a single consensus node is considered a validator, and the costs of proxy nodes required for node operations are excluded. Some networks such as Klaytn require a proxy node to be installed to participate in running a validator, but since this table calculates the cost of operating with a single node, it may differ from the actual operating costs, and the annual reward levels may also vary.

Cosmos Hub and Klaytn, for instance, offer over 10\% APR while other chains offer lower rewards. In terms of upfront investment, hundreds of thousands of dollars of capital should be staked on most chains to expect profits over the costs of running validators.

In the case of Ethereum, the amount that can be staked per validator is fixed (32 ETH), profits exceeding the operating costs can be expected only from running multiple validators on a single client.

\subsection{Network Stability from an Economic Perspective}
It is generally believed that Proof-of-Stake (PoS) blockchains require a higher cost for a double-spend attack than Proof-of-Work (PoW) blockchains due to the higher cost of staking. To commit a double-spend attack on a PoS blockchain, an attacker is expected to acquire over the majority of the staked native tokens, which is largely costly. In addition, a blockchain with a high staking ratio can be considered a relatively secure network, as it becomes more difficult for an attacker to obtain the native tokens needed to launch an attack.

The value of a network can be considered proportional to the circulating amount of its native tokens \cite{18}. This is because the higher the circulation, the more active the network is considered to be. But if a permissionless blockchain has a low staking ratio to the circulating supply, an attacker could potentially attack the network \cite{17} by purchasing a large number of tokens in circulation \cite{12} \cite{19}. For instance, on a Proof-of-Stake blockchain that achieves consensus proportional to the stake, a double-spend attack \cite{23} becomes possible if an attacker acquires 2/3 of the total staked amount needed for consensus \cite{13}. On the other hand, a blockchain network that uses consensus proportional to the number of validators may require less than 2/3 of the total staked amount because consensus can be made when reaching 2/3 of the number of validators sorted by ascending staking amount. Therefore, an attacker may be able to launch an attack at a relatively low cost.

To compare network resistance to attacks, we calculated the required staking ratio to the tradable quantity for consensus by each network as described in [Table 6]. As the definition of tradable quantity slightly differs by network, we have defined it as the quantity that can be secured through normal transactions, which is total supply or circulating supply in here. And the quantity already minted but designated as a reserve or burnt is excluded from the tradable quantity.

\begin{table}[h]
    \centering
    \captionsetup{justification=centering}
    \includegraphics[width=0.5\textwidth]{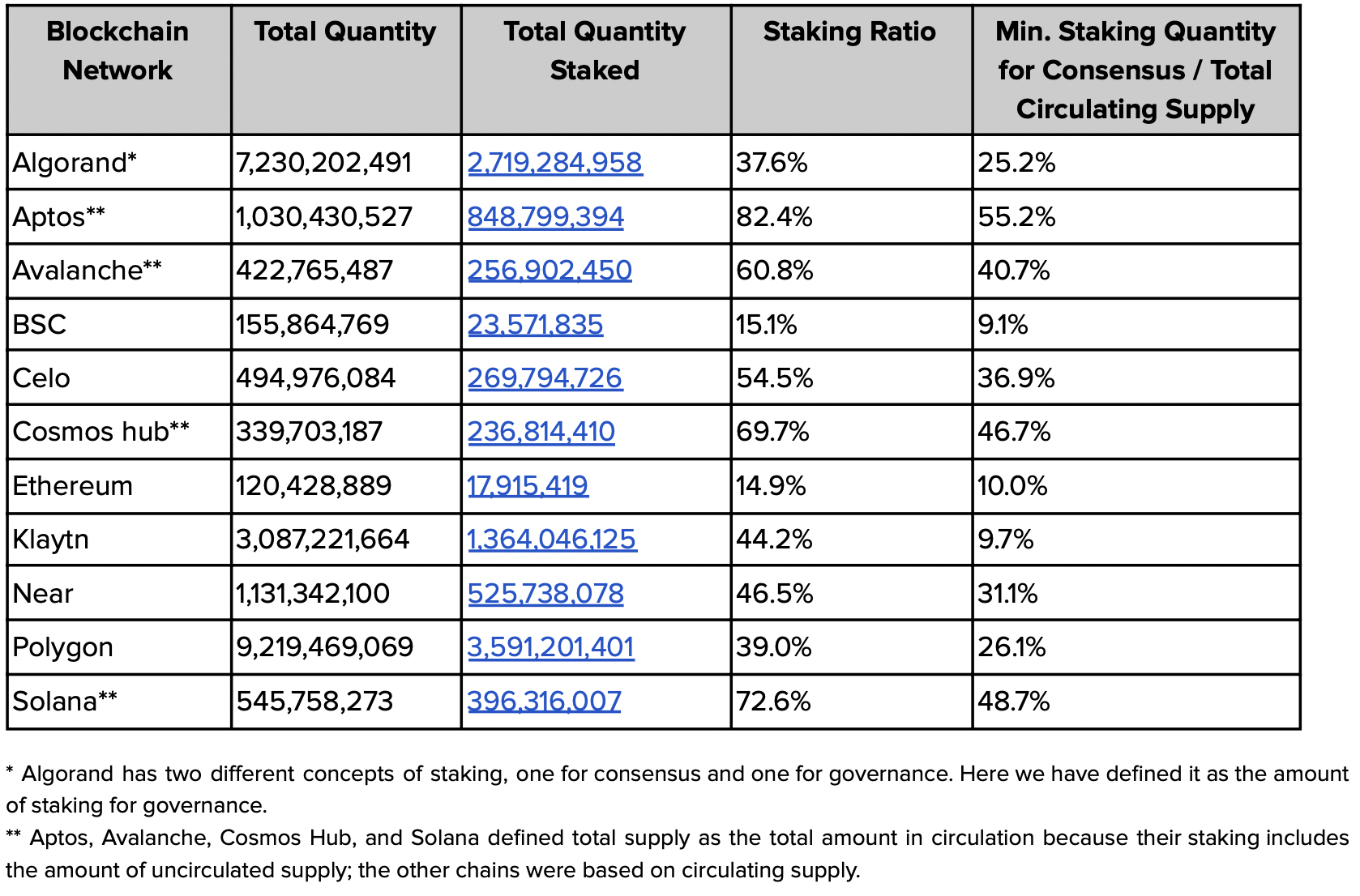}
    \caption{\textit{Staking quantity for consensus out of the total circulating supply as of Apr 21 2023}}
    \label{table:6}
\end{table}

\textit{Table~\ref{table:6}} shows each network’s total circulating supply and the minimum staking amount for consensus divided by the total circulating supply. If a network has a large circulating supply of its native tokens and small staking amount, an attacker could make a large profit from a successful attack. In this case, a double-spend attack can be attempted even at the risk of enormous costs to attack. From this perspective, blockchains with a high staking ratio such as Aptos, Cosmos Hub, and Solana can be considered relatively secure against double-spend attacks.

\begin{figure}[h]
    \centering
    \captionsetup{justification=centering}
    \includegraphics[width=0.5\textwidth]{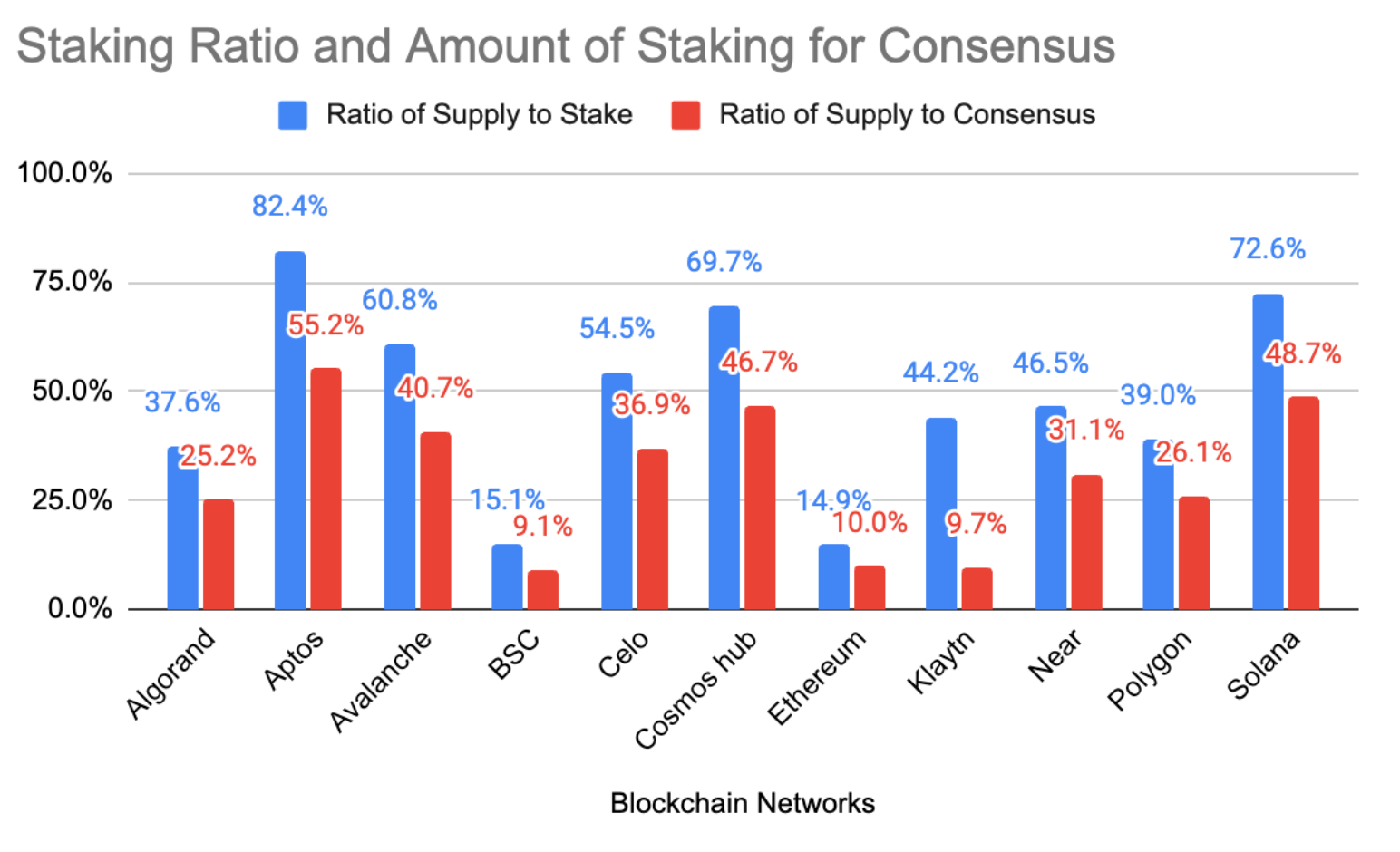}
    \caption{\textit{The staking ratio and amount of staking for consensus of different blockchain networks}}
    \label{fig:4}
\end{figure}

As described in \textit{Fig.~\ref{fig:4}}, on a network with consensus proportional to the number of validators, the amount of native tokens needed for consensus may be less than 2/3 of the amount staked. As Algorand, BNB Chain, and Klaytn need a relatively small quantity of native tokens for consensus, an attacker can attempt an attack to seize a network such as a double-spend attack if acquiring nearly 10\% of the circulating supply. However, committing such an attack would be based on the assumption that the attacker can acquire a sufficient number of validators; otherwise, such an attack would be deemed impossible.

Celo blockchain uses a consensus method proportional to the number of validators but needs a relatively large amount of native tokens for consensus, which means staking amount is relatively evenly distributed among validators. This makes the Celo network relatively resistant to attacks from an economic standpoint.

Moving on, let's analyze how much capital an attacker would actually need to disrupt a blockchain network. This will help estimate how much capital is needed for an attack and assess the resistance of a blockchain network to an attack.

Assuming that the attacker is unable to take the amount staked, the attacker then would need to purchase and stake the number of native tokens from the market for an attack, which would be at least 50\% of the current staking amount according to the proportional consensus method. This is because as soon as the attacker stakes half of the current staking amount, the attacker's stake becomes 1/3 of the total staking amount. Therefore, if we know the price of the native token and the amount staked, one could roughly estimate the amount of capital an attacker would require to attack the target blockchain network.

\begin{table}[h]
    \centering
    \captionsetup{justification=centering}
    \includegraphics[width=0.5\textwidth]{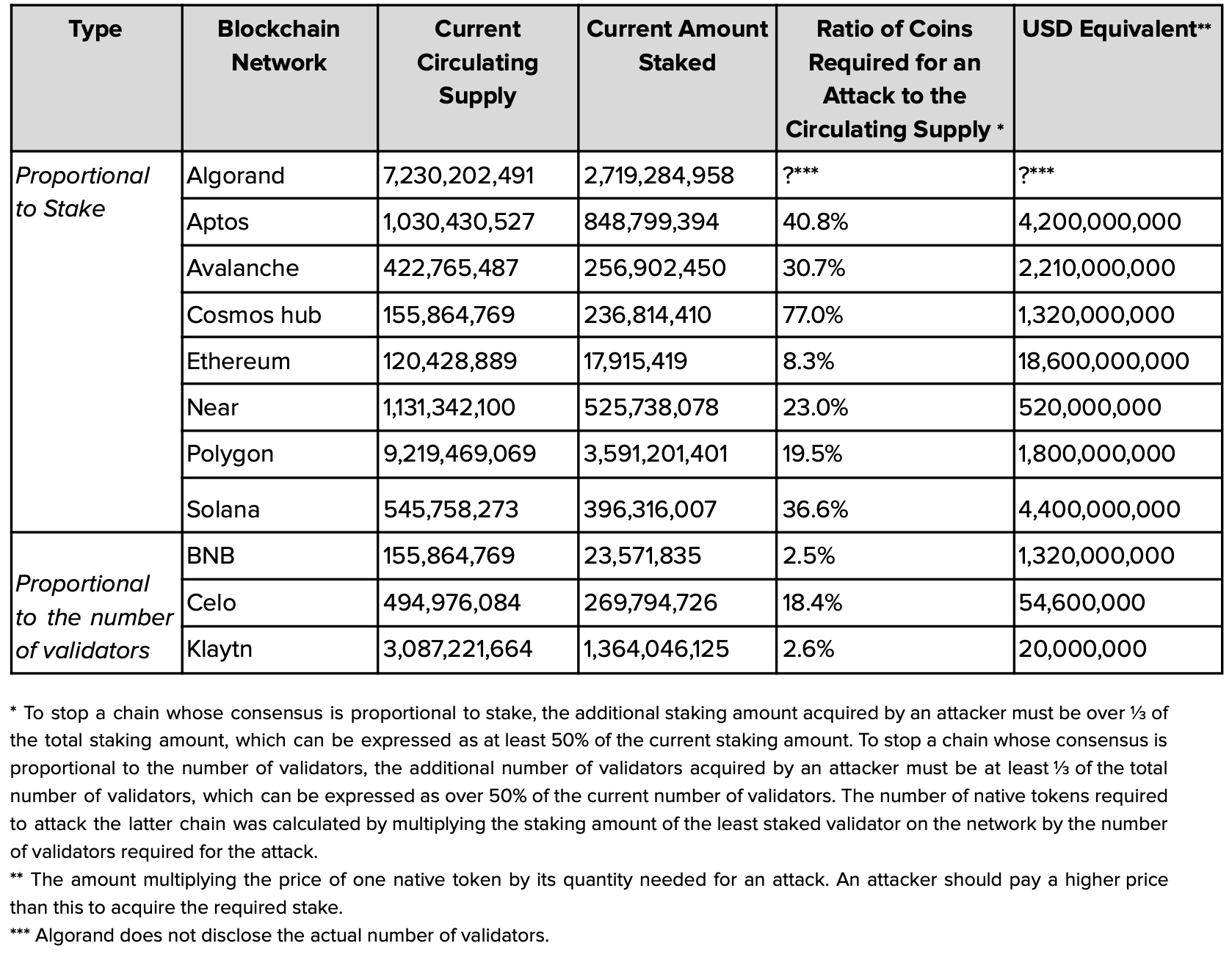}
    \caption{\textit{Quantity of native tokens and capital required to attack a network as of April 21 2023}}
    \label{table:7}
\end{table}

\textit{Table~\ref{table:7}} demonstrates a rough idea of how economically attack-resistant each blockchain network is. It indicates that an attacker would need a very large amount of capital to disrupt a blockchain network which is proportional to stake, which confirms that Proof-of-Stake based blockchain networks have a high level of attack resistance. In the case of Ethereum, the cost of an attack was estimated to be at least \$18 billion, making it the most resistant blockchain among those analyzed.

On the other hand, networks with their consensus mechanism which is proportional to the number of validators have relatively lower level of resistance to security attacks. This is because the quantity required for an attack is proportional to the least staked validator. Therefore, these networks need to either limit the maximum number of validators or increase the minimum staking amount to improve attack resistance, thereby properly maintaining their accessibility to participation, stability, and reliability.

\section{Results of Comparative Analysis on Network Openness Levels}

The actual cost of a network attack can vary depending on a variety of factors, including market conditions and price fluctuations, so the analysis results should be considered a rough indicator.

\begin{table}[h]
    \centering
    \captionsetup{justification=centering}
    \includegraphics[width=0.5\textwidth]{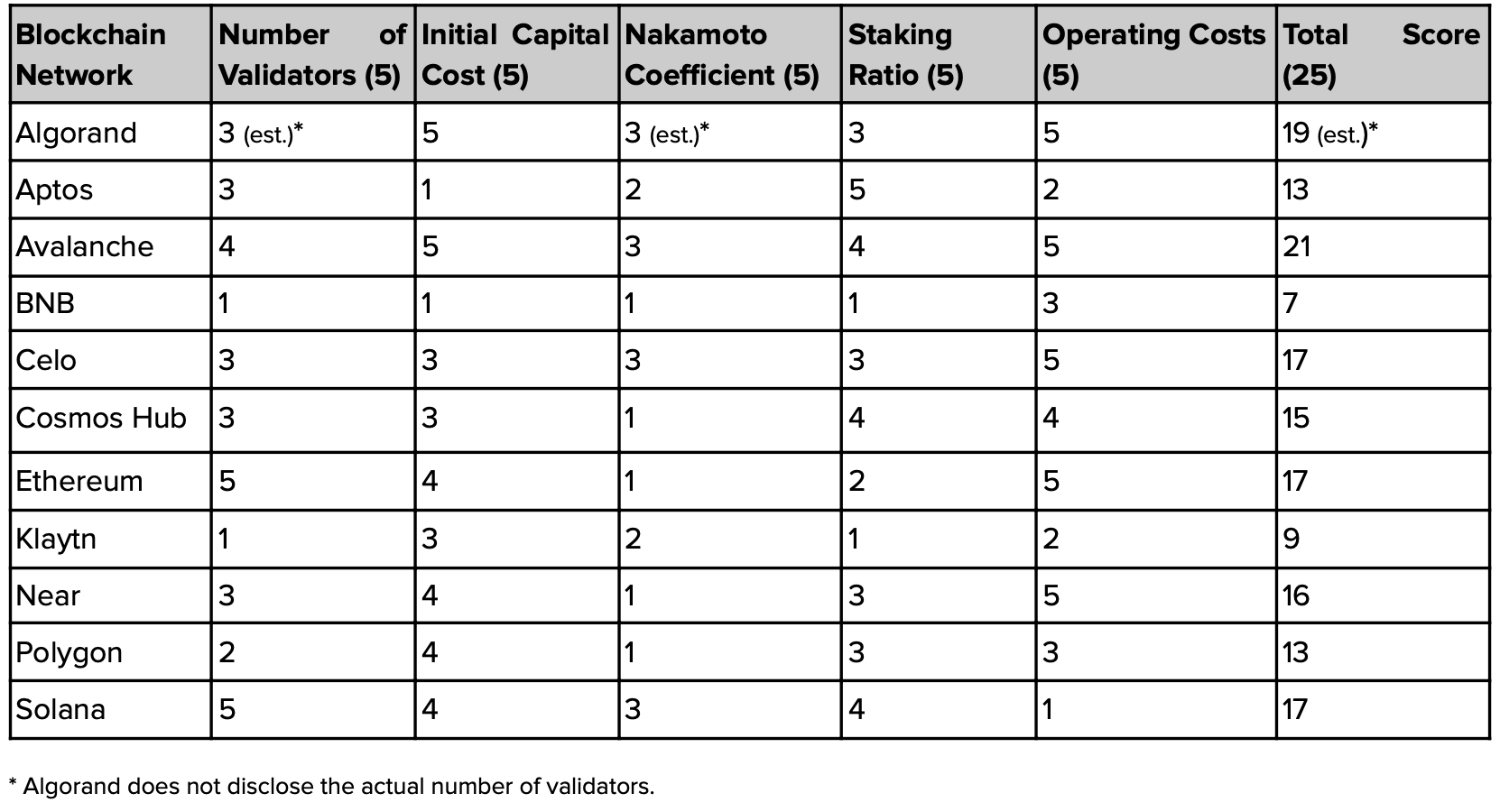}
    \caption{\textit{A relative level of openness among blockchain networks (minimum 1, maximum 5)}}
    \label{table:8}
\end{table}

The scores of the relative openness levels among the eleven blockchain networks analyzed is summarized in \textit{Table~\ref{table:8}}, in which each score closer to 5 indicates a higher level of openness while a score closer to 1 indicates a lower level of openness.

Solana and Avalanche networks were assessed to be highly open, with relatively high scores in the number of validators, initial capital cost, and capital concentration. These factors were combined to contribute to increasing their openness.

As Algorand does not disclose the exact number of validators, the exact result is not available, but it is assumed to be highly open based on a combination of data sources. While Ethereum scored high in several metrics, it scored low in the areas of capital concentration and network stability, which means that there are some limitations to the openness of the Ethereum network.

\begin{figure}[h]
    \centering
    \captionsetup{justification=centering}
    \includegraphics[width=0.5\textwidth]{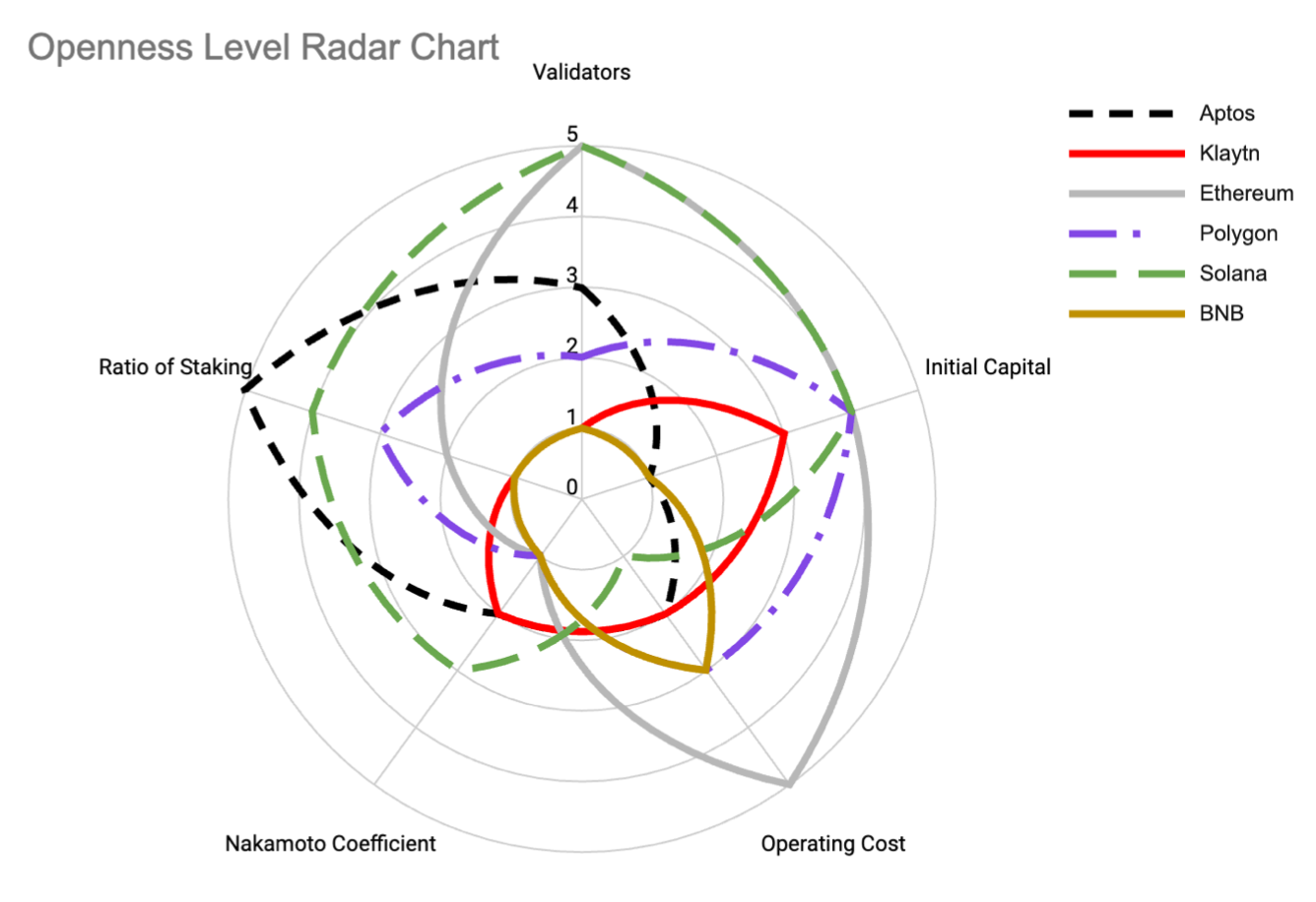}
    \caption{\textit{Openness Level Radar Chart}}
    \label{fig:5}
\end{figure}

Aptos and BNB Chain scored lower in initial capital cost and decentralization, and their total scores were also relatively low. Given the fact that both Klaytn and Polygon blockchains have been permissioned blockchains, their openness scores do not mean much, but if they transition into a permissionless blockchain, they could likely be less open. The radar chart \cite{21} depicted in \textit{Fig.~\ref{fig:5}} visualizes the openness levels of six blockchain networks out of the eleven blockchain networks bench-marked in this study. It is suggested that the relatively highly-open Ethereum and Solana taking up a larger area, while BNB Chain and Klaytn occupy a relatively small area. This chart helps compare and assess the openness levels of the networks at a glance.

\section{Conclusions}
In this comparative analysis, we compared the openness levels of different Proof-of-Stake (PoS) blockchain networks and summarized with the following insightful results. 

\begin{enumerate}
    \item Proof-of-stake networks could be categorized by 2 evaluation methods: proportional to staking and proportional to the number of validators.
    \item Networks with proportional staking have a relatively large number of validators and are relatively highly-stable with high potential for capital concentration.
    \item Networks with consensus proportional to the number of validators are relatively highly-open in terms of capital concentration but could be vulnerable to network attacks if there are large variations in the amount staked.
    \item Some blockchain networks are with high initial capital requirements, making it difficult to participate as a validator without support from the corresponding foundation of the blockchain networks.
    \item High-performance blockchains require high hardware specifications, resulting in relatively high operating costs.
    \item Algorand, Avalanche, Celo, and Solana demonstrate a higher level of network openness. 
    \item Cosmos Hub, Ethereum, and NEAR Protocol are moderately open. 
    \item Ethereum scored very high on many metrics, including the number of validators, initial capital, and operating costs, but scored lower on capital concentration and staking ratio.
    \item The openness of permissionless networks that require an extremely large initial capital is found to be not significantly different from permissioned blockchain networks.
\end{enumerate}

The results summarized in this comparative analysis can help understand how openness and stability of blockchain networks affect each other and could match with which strategy to adopt in order to improve the openness of a permissionless network. Also, so as to transition a permissioned blockchain network into an effectively permissionless blockchain, the following could be taken into account.

\begin{enumerate}
    \item Choose a suitable consensus method for a network between a mechanism of proportional to stake and that of proportional to the number of validators 
    \item Properly set the number of validators, initial capital cost, and operating costs 
    \item Determine the level of capital concentration and staking ratio to circulating supply to maintain network reliability
\end{enumerate}

Designing a methodology to meet the openness level while keeping network reliability based on the above will help the process of converting a permissioned network into a network architecture with a more permissionless setting. For the future research and development, permissioned blockchains such as Klaytn to transition into a network architecture with a more permissionless setting, the key factors are: 1) deciding whether consensus that is proportional to stake or number of validators is more suitable, 2) properly setting the number of validators, initial capital cost, and operating costs, and 3) determining the level of capital concentration and staking ratio to circulating supply to maintain network reliability.

The research team at Klaytn Foundation would hope the findings from this comparative analysis contributing to the blockchain network segment and blockchain industry as a whole when it comes to designing and implementing a permissionless blockchain network and ecosystem.

\bibliographystyle{ieeetr}
\bibliography{fork}
\vskip -2\baselineskip plus -1fil
\begin{IEEEbiography}[{\includegraphics[width=1in,height=1in,clip,keepaspectratio]{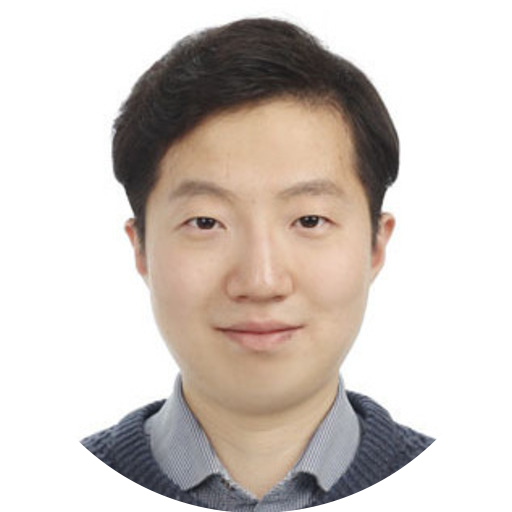}}]{Dr. Jiseong Noh} is currently a Researcher Engineer at Klaytn Foundation and the Research Director at TEEware, Inc. He has been with National Security Research Institute (NSRI) as a Senior Researcher. He has been working on network and system security and blockchain consensus and security. Dr. Noh received his Ph.D. in Computer Science and Information Security from Korea Advanced Institute of Science and Technology (KAIST). 
\end{IEEEbiography}
\vskip -2\baselineskip plus -1fil
\begin{IEEEbiography}[{\includegraphics[width=1in,height=1in,clip,keepaspectratio]{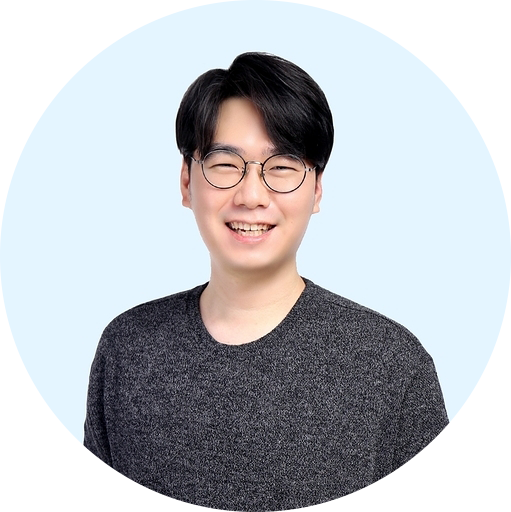}}]{Mr. Donghwan Kwon} is a Security Analyst and Software Developer specialized in blockchain core protocols. He is currently the Lead of core project at Klaytn Foundation. Previously, he tested the security of enterprise systems as a white hacker. He received a master's degree in Information Security from KAIST.
\end{IEEEbiography}
\vskip -2\baselineskip plus -1fil
\begin{IEEEbiography}[{\includegraphics[width=1in,height=1in,clip,keepaspectratio]{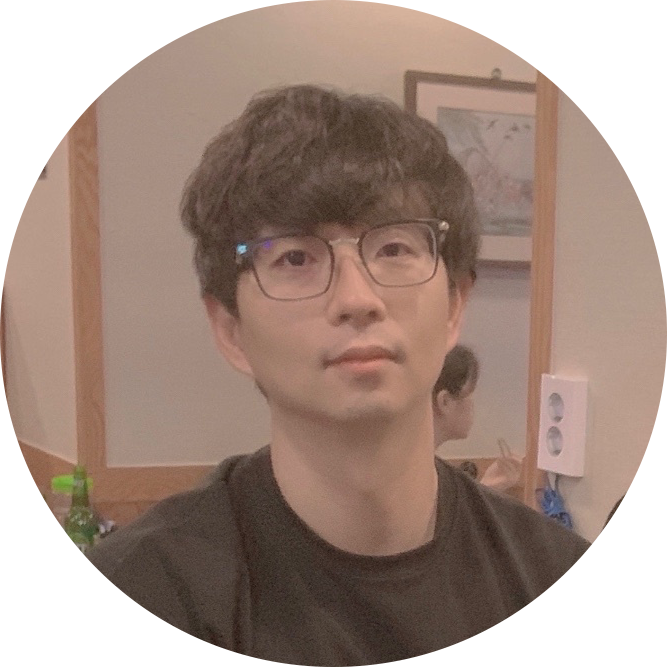}}]{Dr. Soohwan Cho} is currently a Research Engineer at Klaytn Foundation. He previously worked at a blockchain research institute and an operating system company. Dr. Cho received his Ph.D. in Computer Science from Sogang University.
\end{IEEEbiography}
\vskip -2\baselineskip plus -1fil
\begin{IEEEbiography}[{\includegraphics[width=1in,height=1in,clip,keepaspectratio]{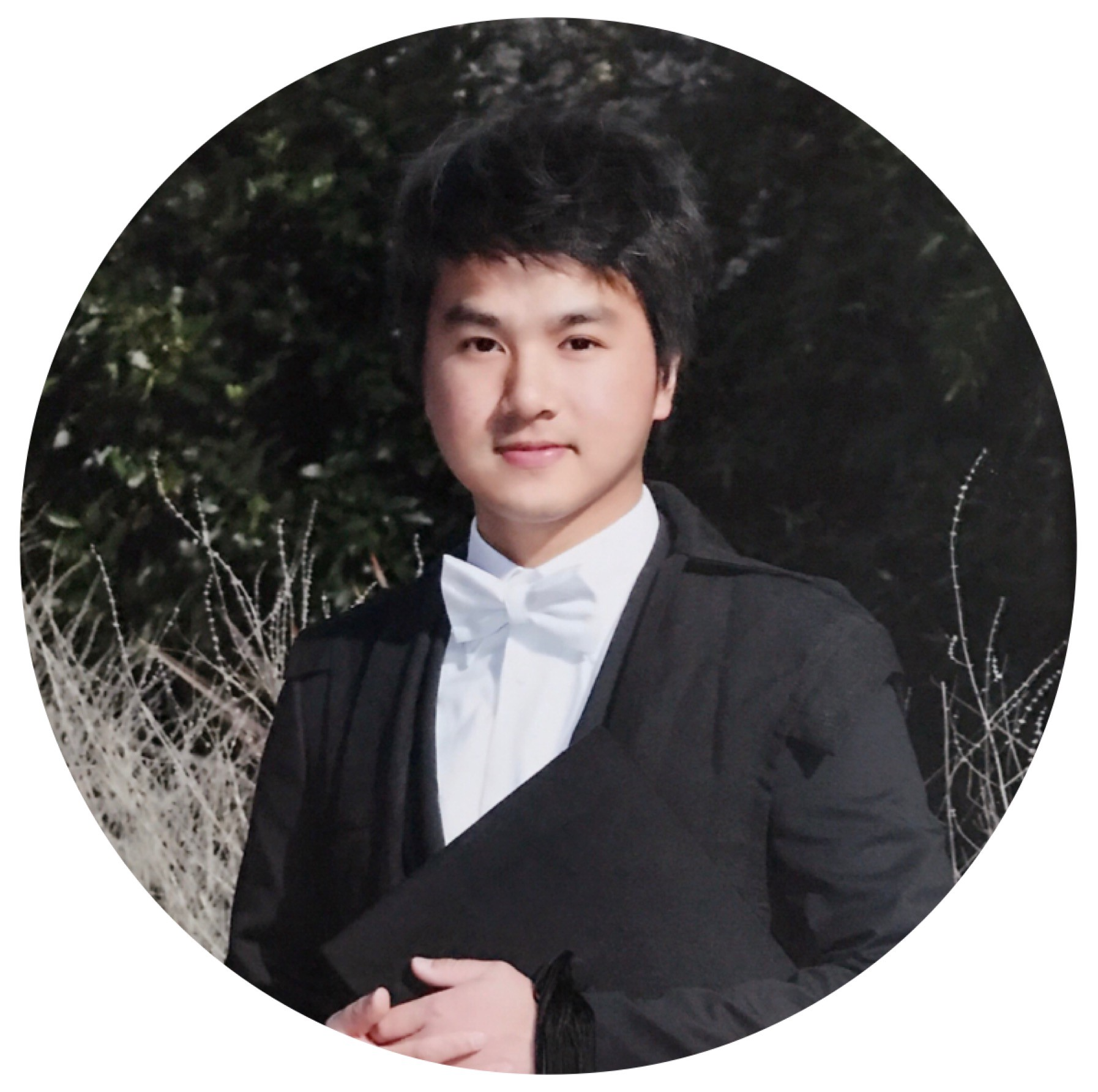}}]{Dr. Neo C.K. Yiu IEEE} is an experienced technology leader focusing on distributed intelligent system R\&Ds. Neo is currently the Head of Technology at Klaytn Foundation. Formerly the Group Blockchain Lead at De Beers Group, Neo has also been the Director of Technology at Oxford Blockchain Society and the Technical Advisor for UCL-CBT and other organizations in the industry. Neo holds a Ph.D. in Engineering (in view) from University of Cambridge and a MSc in Computer Science from University of Oxford.
\end{IEEEbiography}
\vfill
\end{document}